\documentclass[12pt,psfig]{iopart}
\usepackage{iopams}
\usepackage{epsf}
\usepackage{euscript}
\usepackage{graphicx}
\usepackage{setstack} 
\usepackage{amsthm}   
  \theoremstyle{plain}
  \newtheorem{thm}{Theorem}[section]

  \theoremstyle{plain}
  \newtheorem{lemma}[thm]{Lemma}

  \theoremstyle{plain}
  \newtheorem*{lemma*}{Lemma}

  \theoremstyle{definition}
  \newtheorem{defn}[thm]{Definition}

  \theoremstyle{plain}

  \theoremstyle{plain}
  \newtheorem*{cor*}{Corollary}

  \theoremstyle{remark}
  \newtheorem*{rem*}{Remark}

  \theoremstyle{plain}
  
\date{\today}
\eqnobysec   
\newcommand{\Cov}{\mathop{\mathrm{Cov}}\nolimits}
\newcommand{\Var}{\mathop{\mathrm{Var}}\nolimits}

\newcommand{\GS}{\mathcal{GS}_d(\lambda)}

\newcommand{\be}{\begin{eqnarray}}
\newcommand{\ee}{\end{eqnarray}}

\begin{document}

\title{Percolating level sets of the adjacency eigenvectors of $d$-regular graphs}
\author{Yehonatan Elon$ ^1$ and Uzy Smilansky$ ^{1,2}$}
\address {$^1$Department of Physics of Complex Systems,\\ The Weizmann
Institute of Science, 76100 Rehovot, Israel\\
$^2$Cardiff School of Mathematics and WIMCS, Cardiff University,\\
Senghennydd road, Cardiff CF24 4AG, UK.}

\begin{abstract}
One of the most surprising discoveries in quantum chaos was that
nodal domains of eigenfunctions of quantum-chaotic billiards and
maps in the semi-classical limit display critical percolation. Here
we extend these studies to the level sets of the adjacency
eigenvectors of $d$-regular graphs. Numerical computations show that
the statistics of the largest level sets (the maximal connected
components of the graph for which the eigenvector exceeds a
prescribed value) depend critically on the level. The critical level
is a function of the eigenvalue and the degree $d$. To explain the
observed behavior we study  a random Gaussian waves ensemble over
the $d$-regular tree. For this model, we prove the existence of a
critical threshold. Using the local tree property of $d$-regular
graphs, and assuming the (local) applicability of the random waves
model, we can compute the critical percolation level and reproduce
the numerical simulations. These results support the random-waves
model for random regular graphs, suggested in \cite{Elon08} and
provides an extension to Bogomolny's percolation model
\cite{Bogomolny02} for two-dimensional chaotic billiards.
\end{abstract}

\section{Introduction}\label{sec:intro}

The statistics of the adjacency spectrum of random $d$-regular
graphs (in the limit of large vertex number) displays the generic
attributes associated with the spectra of quantum Hamiltonians (in
the limit $\hbar \rightarrow 0$) whose classical dynamics is
chaotic. This observation which was first based on numerical
simulations \cite {Rudnick}, was further substantiated in
\cite{OrenI,OrenII} where a link was made between the distribution
of cycle counts in $d$-regular graphs, spectral statistics and
Random Matrix Theory. In a way, the association of large $d$-
regular graphs with chaotic dynamics is natural:  The time evolution
of random walks on a typical regular graph is known to mix
exponentially fast \cite{Broder,Friedman}. At the same time, the
adjacency operator of a random regular graph is a symmetric matrix
with identically (though not independently) distributed random
variables (see section \ref{subsec:NDgraphs}). Thus, it does not
come as a surprise that in the limit of large graphs with a fixed
degree, the spectral statistics of a typical regular graph should
follow some of the universality classes related to systems with
chaotic (or mixing) dynamics.

Recently,  the study of level sets of eigenfunctions for two
dimensional billiards and in particular the zero level sets (which
are the nodal domains) have gained a considerable attention as
possible indicators for the  dynamics of the underlying classical
system
\cite{Handy,Blum,Bogomolny02,Bogomolny07,BogomolnySle,Keating}. The
distribution of the (normalized) number of nodal domains in two
dimensional  billiards was examined in \cite{Blum}. For chaotic
billiards, this distribution was observed to converge in the semi
classical limit into a universal measure, independent of the
idiosyncratic dynamical characteristics of the billiard. These
findings found an intriguing explanation in \cite{Bogomolny02},
where it was suggested that the distribution of nodal domains can be
approximated by a non-correlated percolation process. Further
connections with percolation theory were discovered in \cite
{BogomolnySle,Keating} who showed that the boundary of the
percolating domain reveal the $SLE_6$ statistics. These observations
motivated the research reported in the present paper, where we
examine the distribution of the adjacency eigenvectors for typical
regular graphs. In particular we investigate the morphology of the
associated \textit{level sets}: Given a graph $G$ and a real
function $f(G)$ which is defined on the vertices of $G$, the
$\alpha$-level sets of $f$ in $G$ are the maximal connected
components of $G$, on which $f$ exceeds the value $\alpha$.

Level sets of a special interest are the zero level sets - the nodal
domains. In a previous paper \cite{Elon08}, the dependence of the
expected number of nodal domains on the spectral parameter was
studied. The numerically observed patterns were accurately
reproduced by assuming a random waves model where the distribution
of the eigenvectors converges to that of a Gaussian random field on
the regular tree $T_d$.

Here we study further the morphology of level sets, and in
particular investigate the percolation transition observed
numerically at a critical level $\alpha_c$ which depends on the
eigenvalue under consideration. Invoking again the random wave
hypothesis we are able to reproduce the numerical simulations, and
thus bring further supporting evidence for the applicability of the
random waves model. In this way we provide the discrete analogue to
Bogomolny's percolation hypothesis \cite{Bogomolny02}, and shed more
light on the surprising connection between percolation and spectral
theory. While for two dimensional billiards, if a critical level set
exist it must be the nodal set by duality arguments (e.g.
\cite{Stauffer}), the fact that in the present case the percolation
threshold typically occurs at $\alpha_c\ne 0$ should not come as a
surprise. This is consistent with classical results about
probabilistic percolation on graphs \cite{Alon04}.

The paper is organized in the following way. After some necessary
preliminaries and a review of some pertinent results, we  describe
the results of numerical simulations where the distribution of the
largest $\alpha$-level set for different eigenvectors of random
regular graphs were studied. The numerically deduced critical level
$\alpha_c(\lambda,d)$ as a function of the eigenvalue and the degree
of the graph summarizes these computation. The next chapter is
dedicated to a systematic construction of a Gaussian random waves
model on the $d$-regular tree  \cite{Elon09}. For this process, we
are able to prove that a percolation transition occurs. We then show
that the critical level sets computed for $T_d$ reproduce accurately
the numerical simulations for $d$ regular graphs, which is
reasonable in light of the local tree property. Yet, a rigorous
proof of this observation is lacking. The paper is concluded by a
comparison between the observed critical behavior and the
percolation model for two dimensional billiards.

\section{Preliminaries}\label{sec:Preliminaries}
\subsection{Elementary definitions}\label{sec:definitions}
 A graph $G$ is a discrete set of
vertices, connected by edges. We denote the size of a graph by
$|G|=n$, where by an abuse of notation, we use the symbol $G$ to
denote both the graph and its set of vertices. We will consider only
simple graphs, i.e. graphs containing no loops or multiple edges.  A
graph $G$ is \textit{d-regular} if for every vertex $v\in G$, the
degree of $v$ (or the number of edges connected to $v$) is exactly
$d$.  For vertices $v,v'\in G$, we define the graph distance
$|v-v'|$ as the length of the shortest walk in $G$ from $v$ to $v'$.

A graph is completely specified by its adjacency operator $A$, where
$A_{ij}=1$ if $v_i,v_j\in G$ are adjacent vertices, or zero
otherwise. As the adjacency operator is real and symmetric, it has
$n$ real eigenvalues. We will denote the spectrum and eigenvectors
of the adjacency operator by
 \be\nonumber
  \lambda_1\le\lambda_2...\le\lambda_n\quad\quad
  (A-\lambda_j)f_j=0
 \ee
The eigenvectors are normalized through the paper according to the
convention
 \be\nonumber
  \sum_{v\in G}f_j^2(v)=n
 \ee
so that $\Var(f(v))$ does not vanish as $n\rightarrow\infty$.
 \subsection{$G(n,d)$ - the ensemble of random regular graphs}\label{subsec:NDgraphs}
For a given value of $n,d\in\mathbb N$, the ensemble $G(n,d)$
consists of all $d$ regular graphs on $n$ vertices,
equipped with the uniform measure.\\
The  geometrical and spectral properties of $G(n,d)$ have been
extensively studied for more than 30 years, and are successfully
applied in various fields such as combinatorics, information theory,
pseudo-randomness and more (see \cite{LinialExpanders} for a
review). In the following we would like to investigate the
properties of the eigenvectors of a typical $(n,d)$ graph, for a
fixed $d\ge3$ in the limit $n\rightarrow\infty$. Here and in the
following, by stating that a typical $(n,d)$ graph has a property
$\mathcal X$, we mean that the probability that a graph $G\in
G(n,d)$ has the property $\mathcal X$ converges to one as
$n\rightarrow\infty$.

In \cite{Wormald} the distribution of short cycles in $G(n,d)$ is
calculated. Denoting by $C_k$ the number of independent $k$-cycles
in a graph, it is shown that the distribution of $\{C_k\},\ 3\le k
le \log_{d-1}n/2 $ converges as $n\rightarrow\infty$ into
independent Poisson random variables with an expectation value
 \be\label{eq:NumCycle}
  \mathbb{E}(C_k)=\frac{(d-1)^k}{2k}\ .
 \ee
Since the expected number of $k$-cycles does not increase with the
size of the graph, we find that a ball of radius $c\log n$ around a
random vertex has a tree structure with probability $1-n^{c-1}$, for
every $c<1/2$. As a result, the local structure of an $(n,d)$ graph
near most of its vertices is identical to that of $T_d$ - the
$d$-regular (infinite) tree.\\
The diameter of an $(n,d)$ graph, i.e. the maximal distance between
vertices in $G$, is given by \cite{Bollobas}
 \be\label{eq:diameter}
  \textrm{diam}(G) = \log(n \log n) + O(1)
 \ee
(here and in what follows the logarithm base is $(d-1)$). This
result shows that the typical distance between vertices along the
boundary of the 'local tree' is of the same magnitude as the
distance between two arbitrary vertices in $G$.

The local resemblance between $(n,d)$ graphs and $T_d$ is reflected
in the spectral density of their adjacency operator as well. Both
the spectrum of the tree \cite{Cartier} and the limiting spectral
density of $G(n,d)$ \cite{McKay} are supported on the interval
 \be\label{eq:TdSpecrum}
  \sigma(T_d)=[-2\sqrt{d-1},2\sqrt{d-1}]
 \ee
with a spectral density, given by
 \be\label{eq:mcKay}
  p(\lambda)=\frac{d}{2\pi}\frac{\sqrt{4(d-1)-\lambda^2}}{d^2-\lambda^2}
 \ee

For every connected regular graph which is not bipartite, the unique
stationary distribution for random walks on the graph is uniform
over the vertices. The rate of convergence into the stationary
distribution is dictated by $P_{v,v'}^{(k)}$ - the probability that
a walk of length $k$ which begins at the vertex $v$ will terminate
in $v'$. In \cite{Friedman} it is shown that for a typical $(n,d)$
graph
 \be\label{eq:ConvergenceRW}
  \limsup_{k\rightarrow\infty}\left|P_{v,v'}^{(k)}-\frac1n\right|^{1/k}=e^{-\gamma}
 \ee
where the Lyapunov exponent
 \be\nonumber
  \gamma=1-\frac{2\sqrt{d-1}}{d}+O\left(\frac1{\log n}\right)
 \ee
is strictly positive for $d\ge3$. As a result, random walks on a
typical $(n,d)$ graph are exponentially mixing with a Lyapunov
exponent $\gamma\approx1-2\sqrt{d-1}/d$. This observation justifies
attributing the title ``chaotic" to  $(n,d)$ graphs, as was done in
the introduction section. For further detail on expanding
(equivalently, mixing) graphs and the relations between their
spectral, geometrical and dynamical properties we refer the reader
to \cite{LinialExpanders}.
 \subsection{The random waves model for the adjacency eigenvectors of $G\in G(n,d)$}
The main tool in the present study is the random wave model
\cite{Elon08,Elon09}. It is based on the observation that for the
$d$ regular tree $T_d$, the distribution of a typical eigenvector
can be approximated by a real Gaussian process $\GS$. The process
associates random functions $\psi_\omega:T_d\rightarrow\mathbb R$ to
the regular tree, so that for every subset of vertices
$\{v_j\}\subset T_d$, the distribution of any linear combination of
$\psi_\omega(v_j)$ is Gaussian.

A Gaussian process is uniquely characterized by its mean and
covariance operator. Introducing the Chebyshev Polynomials of the
second kind
 \be
  U_k(x)=\frac{\sin\left((k+1)\cos^{-1}(x)\right)}{\sin\left(\cos^{-1}(x)\right)}
 \ee
and the polynomials \cite{Cartier,Brooks}
 \be\label{eq:CofE}
\phi^{(\lambda)}(k)=(d-1)^{-k/2}\left(\frac{d-1}{d}U_k\left(
\textstyle{\frac{\lambda}{2\sqrt{d-1}}}\right)-\frac1dU_{k-2}
\left(\textstyle{\frac{\lambda}{2\sqrt{d-1}}}\right)\right)
\ee It was shown in \cite{Elon08,Elon09} that for every
$\lambda\in\sigma(T_d)$ (\ref{eq:TdSpecrum}), the random Gaussian
process $\GS$ which is characterized by the covariance
 \be\label{eq:CovGS}
  \Cov(\psi_\omega(v),\psi_\omega(v'))=\phi^{(\lambda)}(|v-v'|)
 \ee
Has the following properties:
 \begin{enumerate}
  \item $(A_{T_d}-\lambda I)\psi_\omega=0$ for almost every $\psi_\omega\in\GS$.
  \item $\GS$ is invariant with respect to the symmetries of $T_d$.
  \item The process is normalized, so that $\Var(\psi_\omega(v))=1$.
 \end{enumerate}

The random wave model is based on the conjecture that the
distribution of a typical adjacency eigenvector of a graph $G\in
G(n,d)$ graph, with an eigenvalue $\lambda\in\sigma(T_d)$ converges
locally to that of $\GS$. This hypothesis, was supported by various
numerical tests, and found a partial formal justification in
\cite{Elon09}.  The random waves hypothesis will be the basis for
the analysis of the morphology of level sets in $(n,d)$ graphs,
which will be carried out in this paper. It is the analogue of
Berry's model for the distribution of eigenfunctions for chaotic
billiards \cite{Berry77}.

\section{Level sets percolation on $(n,d)$ graphs}\label{sec:Numerics}
In this section we shall present the numerical evidence which led us
to propose that level sets undergo a percolation transition in the
limit $n\rightarrow \infty$.

For a graph $G$, a real function $f(G)$ and a given
$\alpha\in\mathbb R$, we denote by
 \be\label{eq:IndG}
  \tilde G_\alpha(f)=\{v\in G,f(v)>\alpha\}
 \ee
the induced graph, which is obtained by deleting all the vertices
for which $f(v)$ is below the threshold $\alpha$; The $\alpha$-level
sets of $f$ in $G$ are the connected components of $\tilde
G_\alpha(f)$.\\

\begin{figure}[h]
  \centering
 \scalebox{0.7}{\includegraphics{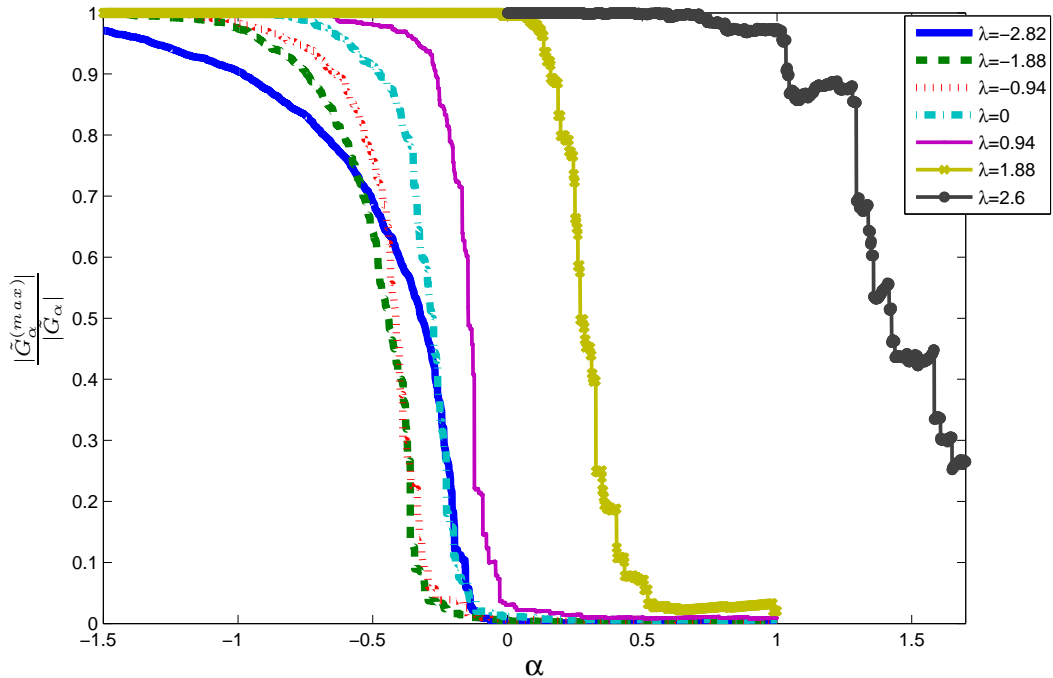}}  
  \caption{The ratio between the magnitude of the largest level set
  to the size of the induced graph $\tilde G_\alpha$ for a single
  realization of a $(4000,3)$ graph. Each curve
  corresponds to one eigenvector, whose level sets are interrogated by increasing $\alpha$.}
    \label{fig:PTNDlambda}
\end{figure}

Since there is no known analytical expression for the distribution
of eigenvectors in an $(n,d)$ graph, we cannot offer an expression
for the limiting distribution of the $\alpha$-level sets for this
ensemble. However, motivated by the resemblance between the spectral
and eigenfunctions statistics for $G(n,d)$ and chaotic billiards, we
have looked for a
numerical evidence to a phase transition in $(n,d)$ level sets.\\
For a given graph $G\in G(n,d)$, $\alpha\in\mathbb{R}$ and an
eigenvector $(A-\lambda I)f=0$, we define $\tilde G_\alpha^{(max)}$
to be the largest component of $\tilde G_\alpha(f)$ (\ref{eq:IndG})
and evaluate the ratio $|\tilde G_\alpha^{(max)}|/|\tilde
G_\alpha|$.\\
We have generated (following \cite{Steger}) and diagonalized (using
MatLab) random regular graphs on up to $4000$ vertices with degrees
ranging from $3$ to $15$. For each eigenvector $f$ of each graph we
have measured the ratio $|\tilde G_\alpha^{(max)}|/|\tilde
G_\alpha|$ while varying $\alpha$ from $\min(f)$ to $\max(f)$. The
ratios $|\tilde G_\alpha|/|\tilde G_\alpha^{(max)}|$ are plotted in
figure \ref{fig:PTNDlambda}as a function of $\alpha$ for a
$3$-regular graph on $4000$ vertices. The different lines
correspond to different values of $\lambda$ are given in the inset.\\
A sharp transition in the normalized size of the level sets is
evident: for every value of $\lambda$ there is a narrow window in
the vicinity of some $\alpha_c(\lambda,d)$, so that for
$\alpha>\alpha_c$, the ratio $|\tilde G_\alpha^{(max)}|/|\tilde
G_\alpha|$ is close to zero, while for $\alpha<\alpha_c$, $|\tilde
G_\alpha^{(max)}|/|\tilde G_\alpha|$ is of order one. The described
phenomenon was observed for all the tested values of $d$
and for all the examined eigenvectors.\\
Moreover, repeating the experiment, while varying the size of the
graph, we have observed that the value of $\alpha_c(\lambda,d)$ does
not vary with $n$, while the transition becomes sharper as $n$
increase.
 \begin{figure}[h]
  \centering
 \scalebox{0.6}{\includegraphics{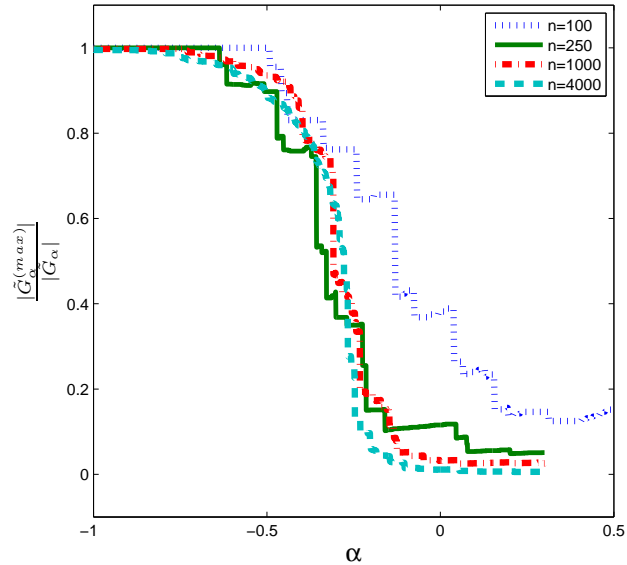}}
  \caption{A comparison of the curve $|\tilde G_\alpha^{(max)}|/|\tilde
  G_\alpha|$, for several $3$-regular graphs of varying size. the curves
  correspond to $n=100,250,1000,4000$ where in each graph we
  consider the eigenvector with the closest eigenvalue to zero.}
    \label{fig:PTNDwindow}
\end{figure}
As an example, in figure \ref{fig:PTNDwindow} we plot the variation
in $|\tilde G_\alpha^{(max)}|/|\tilde G_\alpha|$ for the
eigenvectors which correspond to $\lambda\approx0$, for different
realizations of $3$-regular graphs of various sizes.

This extensive corpus of numerical data provides strong evidence
supporting the existence of a phase transition for the level sets of
$(n,d)$ eigenvectors. Namely, for every $d\ge3$ and
$\lambda\in\sigma(T_d)$, there may exist an
$\alpha_c(\lambda,d)\in\mathbb R$, so that the level sets of a
typical $(n,d)$ eigenvector which corresponds to the eigenvalue
$\lambda$ are all microscopic for $\alpha>\alpha_c$, while for
$\alpha<\alpha_c$ a macroscopic component is expected to appear.

\begin{figure}[h]
  \centering
   \scalebox{0.7}{\includegraphics{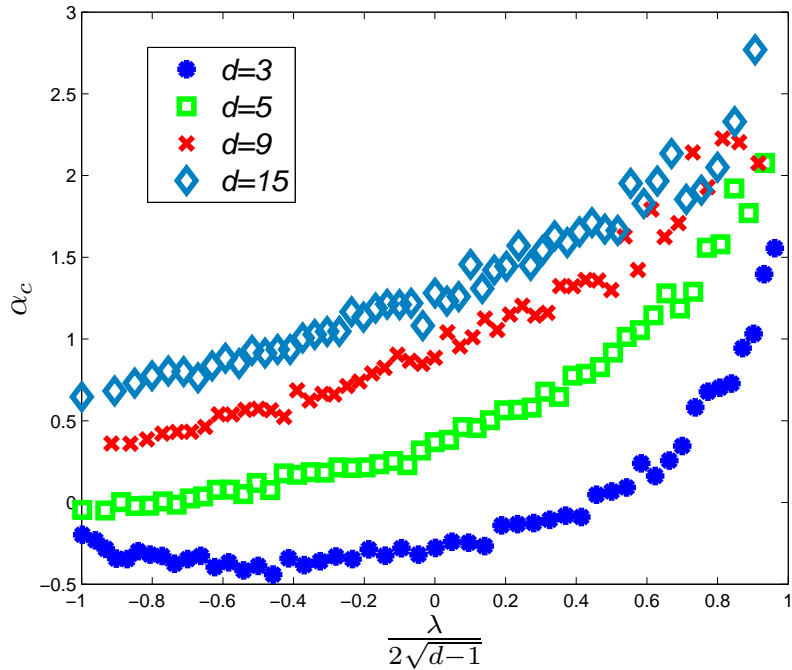}}
   \caption{A numerical estimation to $\alpha_c(\lambda,d)$
    for $d=3,5,6,12$.}
   \label{fig:PTNDalpha_c0}
 \end{figure}

Note that the suggested transition differs from the percolation
hypothesis for chaotic billiards in two main aspects. First, while
for billiards the transition is expected to follow the
characteristics of non-correlated percolation, for $(n,d)$ graphs we
expect correlations to be relevant (as will be discussed in section
\ref{sec:Discussion1}). Second, Unlike the percolation model for
billiards, the critical threshold for regular graphs depends on the
corresponding eigenvalue.\\
In order to estimate the dependence of $\alpha_c(\lambda,d)$ on its
arguments, we have chosen (somewhat arbitrarily) to identify
$\alpha_c(\lambda,d)$ with the steepest point of the curve $|\tilde
G_\alpha^{(max)}|/|\tilde G_\alpha|$, obtained for a graph of size
$n=4000$. In figure \ref{fig:PTNDalpha_c0} we present our numerical
estimate of the critical curves.
The results suggest that for a given value of $d$, the critical
threshold increases monotonically with $\lambda$ for all $d>3$. For
the lowest degree, $d=3$, the critical curve $\alpha_c(\lambda,d=3)$
shows a minimum in the vicinity of $\lambda\approx-0.52$.
\\
In the next sections we shall show that the numerical resulted
summarized above can be reproduced theoretically by considering
first the level sets in the random process $\GS$ on $T_d$, and then
\emph{assuming} the validity of the random waves conjecture for the
adjacency eigenvectors in $G(n,d)$.

 \section{The distribution of level sets in $\GS$}\label{sec:RandomWaves}
In this section we shall study the distribution of the level sets in
$GS$ and prove that they undergo a percolation transition, for which
the critical threshold $\alpha_c(\lambda,d)$ can be computed.

As the process $\GS$ is Gaussian and characterized by the covariance
operator (\ref{eq:CovGS}), it is possible to rigorously analyze its
level sets statistics. Setting
 \be\label{eq:TOmega}
  T_\omega(\alpha)=\{v\in T_d, \psi_\omega(v)>\alpha\}
 \ee
to denote the induced $\alpha$ level sets tree for a given
$\psi_\omega\in\GS$, we show that:
 \begin{thm}\label{Thm:PT}
  $\forall\lambda\in\sigma(T_d)$, there exists an $\alpha_c\in\mathbb{R}$ so
  that for almost every realization $\psi_\omega\in\GS$, $T_\omega(\alpha)$
  has an infinite component for $\alpha<\alpha_c$, but only finite components for
  $\alpha>\alpha_c$.
 \end{thm}
As the proof of the theorem is rather technical, we refer the
interested reader to \cite{Elon09} where  a complete and detailed
proof of the theorem can be found. Here, we shall provide the main
line of the proof, skipping much of the technical aspects.

It is important to note that for a given $\alpha\in\mathbb R$, the
$\alpha$-level sets of the process $\GS$ is a homogeneous vertex
process on $T_d$. That is, the probability measure of the process is
invariant with respect to the symmetries of $T_d$.

For any homogenous process, the probability  that $v,v'\in T_d$
belongs to the same connected component depends only on the distance
between the vertices and will be denoted by $P_{|v-v'|}$. Denoting
the sphere of radius $k$ around $v\in T_d$ by
 \be\label{eq:LamK}
  \Lambda_k(v)=\{v'\in T_d, |v-v'|=k\}
 \ee
we find that the probability that $v$ is connected to its $k$-sphere
is at most $|\Lambda_k(v)|P_k$. Since $|\Lambda_k|=d(d-1)^k$, we
obtain that if
 \be\label{eq:subcritical}
  \limsup_{k\rightarrow\infty}P_k^{1/k}<\frac1{d-1}
 \ee
then the probability that the connected component of $v$ exceeds the
radius $k$ decays exponentially with $k$, so that the probability to
find an infinite component is zero and the process is subcritical.\\
We should note that the opposite statement is not necessarily
correct, i.e. there are percolation processes on $T_d$ for which
$\liminf_{k\rightarrow\infty}P_k^{1/k}>1/(d-1)$, but do not contain
infinite components, due to long range correlations.
 \begin{figure}[h]
  \centering
 \scalebox{0.4}{\includegraphics{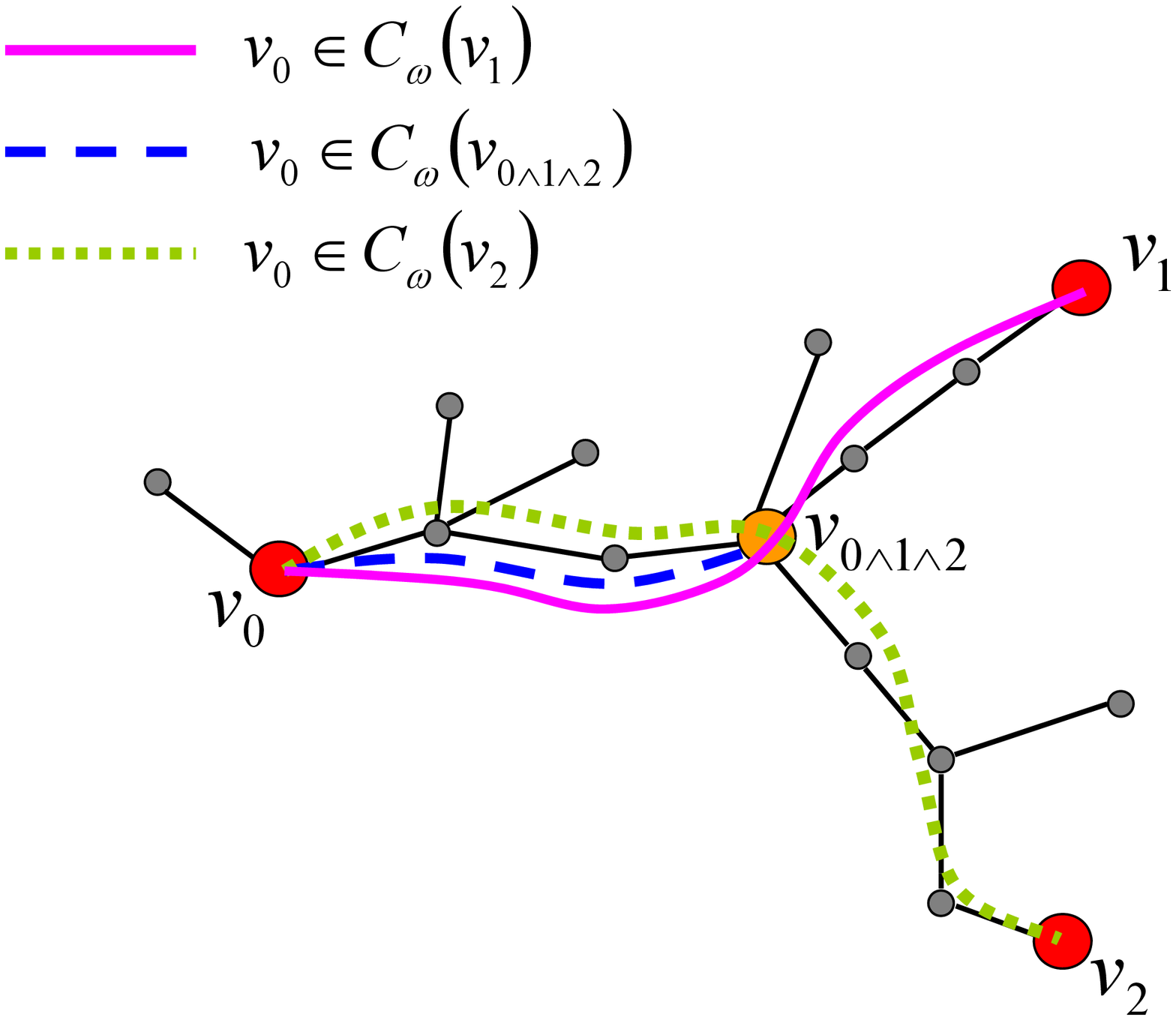}}
  \caption{\footnotesize For a quasi-bernoulli process, the probability to find $v_0\in C_\omega^{\alpha}(v_1)$
   (i.e. that the continuous purple line is entirely occupied in
   $\Gamma_\omega$) conditioned that $v_0\in C_\omega^{\alpha}(v_2)$
   (dotted green line is occupied) is uniformly bounded by the
   probability that $v_0\in C_\omega^{\alpha}(v_1)$, conditioned that
   $v_0\in C_\omega^{\alpha}(v_0\wedge v_1\wedge v_0)$ (dashed blue
   line is occupied). Therefore the effect of values which are obtained
   by the process along the path $v_{0\wedge1\wedge2}\rightarrow v_2$
   has only a limited effect on the path
   $v_{0\wedge1\wedge2}\rightarrow v_1$.}
    \label{fig:QBP}
\end{figure}

The influence of the correlations which are induced by the process
can be formally evaluated according to the next classification of
random processes on trees, introduced in \cite{Lyons}.
 \begin{defn}\label{defn:quasi_bernoulli}
  A random percolation process on a tree graph $\Gamma$, associating
  for all $\omega\in\Omega$ an induced subgraph $\Gamma_\omega\subset\Gamma$,
  is a \textit{quasi Bernoulli} process, if $\exists M<\infty$,
  such that $\forall v_0,v_1,v_2\in\Gamma$:
   \be\label{eq:quasi}
    \frac{\mathbb{P}\left(v_1\in C_{\Gamma_\omega}(v_0)|v_2\in C_{\Gamma_\omega}(v_0)\right)}
    {\mathbb{P}\left(v_1\in C_{\Gamma_\omega}(v_0)|v_{0\wedge1\wedge2}\in C_{\Gamma_\omega}(v_0)\right)}\le M \ .
   \ee
  where $v_{0\wedge1\wedge2}$ is the intersection of the simple paths
  in $\Gamma$ between the three vertices (see figure \ref{fig:QBP})
  and $C_{\Gamma_\omega}(v_0)$ is the connected component of $v_0$ in
  $\Gamma_\omega$.
 \end{defn}
According to definition \ref{defn:quasi_bernoulli}, in a quasi
Bernoulli process, the distribution of the process along any simple
path is only moderately affected by the realization of the process
outside the path. As an example, it can be verified that a Markovian
process is quasi Bernoulli.\\
The quasi Bernoulli classification provides a simple criterion for
the relevance of the long range correlations of the process on its
macroscopic properties, as suggested by the next lemma
\cite{Lyons}\footnote{The theorem as it appears in \cite{Lyons}
characterizes general quasi-Bernoulli process on an arbitrary tree
graph. For the sake of clarity, we consider here only the
restriction of the theorem to invariant percolation processes on
$T_d$.}:
 \begin{lemma}\label{lem:quasi} \textbf{(Lyons)}
  Let $\{\Omega,\mathbb{P}\}$ be an invariant quasi Bernoulli
  process on $T_d$, which associates $\forall\omega\in\Omega$
an induced graph $T_\omega\subset T_d$. If \footnote{Note that for a
quasi Bernoulli process, the limit
$\lim_{|v'-v|\rightarrow\infty}\left(\mathbb{P}(v'\in
C_{T_\omega}(v))\right)^{1/|v-v'|}$ does exists. This can be
verified by restricting the condition (\ref{eq:quasi}) to cases in
which $v_0$ is on the simple path between $v_1$ to $v_2$, i.e.
$v_0=v_{0\wedge1\wedge2}$.}
   \be\label{eq:QBPb1}
    \lim_{k\rightarrow\infty}P_k^{1/k}<\frac1{d-1}
   \ee
  then, all the connected components of $T_\omega$ are almost surely
  finite. If
   \be\label{eq:QBPb2}
    \lim_{k\rightarrow\infty}P_k^{1/k}>\frac1{d-1}
   \ee
  $T_\omega$ will almost surely have a component of an infinite cardinality.
  \phantom{move a little b}$\square$
 \end{lemma}
An equivalent phrasing of the lemma is the following: for a quasi
Bernoulli process on $T_d$, all the components are almost surely
finite if their expected cardinality is finite, while if the
expected cardinality diverges, infinite components will almost
surely exist.\\
As the $\alpha$-level sets are monotonically decreasing in $\alpha$,
we find that in order to prove theorem \ref{Thm:PT} it is enough to
show that
 \begin{itemize}
  \item
   For every $\lambda\in\sigma(T_d)$ and $\alpha\in\mathbb R$, the
   $\alpha$-level sets of $\GS$ are quasi Bernoulli
  \item
   For small enough values of $\alpha$ there almost surely exist
   infinite level sets, while for large enough values the
   $\alpha$-level sets are all finite with probability one.
 \end{itemize}
The first condition guaranties that for every
$\lambda\in\sigma(T_d)$ and $\alpha\in\mathbb R$, the probability to
find an infinite component is either zero or one (according to lemma
\ref{lem:quasi}). Assuming the level sets are quasi Bernoulli, and
since $\lim_{k\rightarrow\infty}P_k^{1/k}(\alpha)$ is monotone in
$\alpha$, the second condition ensures that for every
$\lambda\in\mathbb R$ a supercritical and a subcritical phases
exist, where the transition between the two occurs at
$\alpha_c(\lambda,d)$ which is given by the implicit expression
 \be\label{eq:alpha_c}
  \lim_{k\rightarrow\infty}P_k^{1/k}(\alpha_c)=\frac1{d-1}
 \ee

To verify the existence of a subcritical regime, we note that if a
simple path $U$ of length $k$ is contained in an $\alpha$-level set,
then necessarily $\forall v_j\in U, \psi_\omega(v_j)>\alpha$.
Therefore, setting $\Psi_\omega(U)=\sum_{j=1}^k\psi_\omega(v_j)$, we
obtain that
 \be
  P_k(\alpha)&=&\mathbb P(\forall v_j\in
  U,\psi_\omega(v_j)>\alpha)\\\nonumber
  &<&\mathbb{P}\left(\Psi_\omega(U)>k\alpha\right)
 \ee
Note that $\Psi_\omega(U)$ is a Gaussian random variable, with mean
zero and variance
 \be
  \Var(\Psi_\omega(U))&=&\mathbb{E}\left[\left(\sum_{ij}\psi_\omega(v_i)\psi_\omega(v_j)\right)^2\right]\\\nonumber
  &=&k\left(\phi^{(\lambda)}(0)+2\sum_{j=1}^{k-1}\frac{k-j}k\phi^{(\lambda)}(j)\right)
  <k\Phi^{(\lambda)}
 \ee
where
$\Phi^{(\lambda)}=\phi^{(\lambda)}(0)+2\sum_{j=1}^\infty|\phi^{(\lambda)}(j)|$
(\ref{eq:CofE}).\\
Since $\phi^{(\lambda)}(j)$ is exponentially decreasing in $j$
(\ref{eq:CofE}), we find that $\Phi^{(\lambda)}<\infty$. Therefore
 \be\label{eq:PnBound}
  P_k(\alpha)<\frac1{\sqrt{2\pi k\Phi^{(\lambda)}}}\int_{k\alpha}^\infty\exp\left(-\frac{x^2}{2k\Phi^{(\lambda)}}\right)
  <e^{-\beta\alpha^2k}
 \ee
where $\beta=(2\Phi^{(\lambda)})^{-1}$.\\
As $|\Lambda_k|=d(d-1)^{k-1}$, we obtain from (\ref{eq:subcritical})
and (\ref{eq:PnBound}) that the $\alpha$-level sets are subcritical
for $\alpha>\sqrt{\ln(d-1)/\beta}$.

The existence of infinite $\alpha$-level sets for small enough
$\alpha$ is a straight-forward consequence of \cite{Haggstrom97},
where it is shown that every homogenous vertex percolation process
on $T_d$, for which the survival probability exceeds $d/2(d-1)$ is
supercritical. As $\psi_\omega(v)$ distributes as a normal variable
with mean zero and variance one, the vertex survival probability for
$T_\alpha(\psi_\omega)$ is given by
 \be\label{eq:P0}
  P_0(\alpha)=\frac1{\sqrt{2\pi}}\int_\alpha^\infty e^{-x^2/2}\ .
 \ee
Since $d/2(d-1)\le3/4$ for every $d\ge3$ and as according to
(\ref{eq:P0}) $P_0(\alpha)>3/4$ for $\alpha\le-0.68$, we obtain that
for every $d\ge3$ and $\lambda\in\sigma(T_d)$, the $\alpha$-level
sets of $\GS$ are supercritical for $\alpha<-0.68$.

In order to prove that the $\alpha$-level sets of $\GS$ are quasi
Bernoulli, we have to show that the long range correlations do not
dominate the structure of the random tree $T_\omega(\alpha)$
(\ref{eq:TOmega}). A major step toward this goal is the next
theorem, which identify the following Markov property of the
underlying process $\GS$:
 \begin{figure}[h]
  \centering
 \scalebox{0.6}{\includegraphics{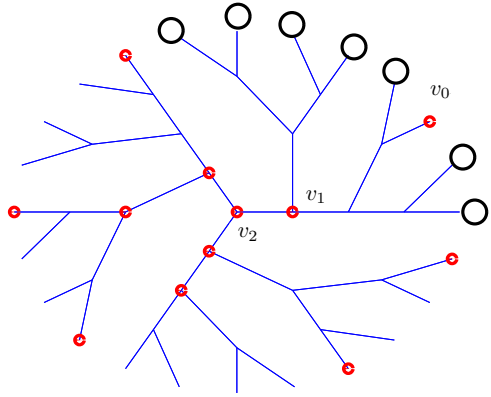}}
  \caption{A set of vertices which corresponds
to the requirements of theorem \ref{thm:Markov}. $v_0,v_1$ and $v_2$
are denoted explicitly; $\{v_j\}_{j\ge3}$ are marked by full red
circles; $\{v'\in\Lambda_k,|v'-v_1|<|v'-v_2|\}$ are marked by hollow
black circles (see \ref{eq:ForCanopy}) for the case $k=3$.}
    \label{fig:PTNDbethe}
\end{figure}
 \begin{thm}\label{thm:Markov}
Let $\{v_j\}_{j=0}^k\subset T_d$ (for $k\ge3$), so that
$|v_1-v_2|=1$, and $v_1$ is on the simple path between $v_0$ to
$\{v_j\}_{j=2}^k$ (see figure \ref{fig:PTNDbethe}). Then
$\forall\lambda\in\sigma(T_d)$, the distribution of
$\psi_\omega(v_0)$ conditioned on $\{\psi_\omega(v_j)\}_{j=1}^k$ is
independent of $\{\psi_\omega(v_j)\}_{j=3}^k$
 \end{thm}
The proof of theorem \ref{thm:Markov} relies on the observation that
for any adjacency eigenvector $\psi:T_d\rightarrow\mathbb R$, the
value of $\psi(v_1),\psi(v_2)$ determines the value of the sum
 \be\label{eq:ForCanopy}
  \sum_{v'\in\Lambda_k(v_1),|v'-v_1|<|v'-v_2|}\psi(v')
 \ee
(see figure \ref{fig:PTNDbethe}). Since the distribution of $\GS$ is
invariant with respect to the symmetries of $T_d$, we obtain from
the last observation that given $\{\psi_\omega(v_j)\}_{j=1}^k$, the
expected value of $\psi_\omega(v_0)$ is determined by
$\psi_\omega(v_1),\psi_\omega(v_2)$.\\
Since non-correlated components of a multi-normal random vector are
also independent (e.g. \cite{Ibragimov}), the theorem follows.
\phantom{move a little b}$\square$

The described dominance of short range correlations in the process
$\GS$, is utilized in \cite{Elon09}, to bound rigorously the effect
of long range correlation in $T_\omega(\alpha)$, resulting in the
identification of $T_\omega(\alpha)$ as a quasi Bernoulli process
and in the establishment of theorem \ref{Thm:PT}.

\begin{figure}[h]
  \centering
 \scalebox{0.8}{\includegraphics{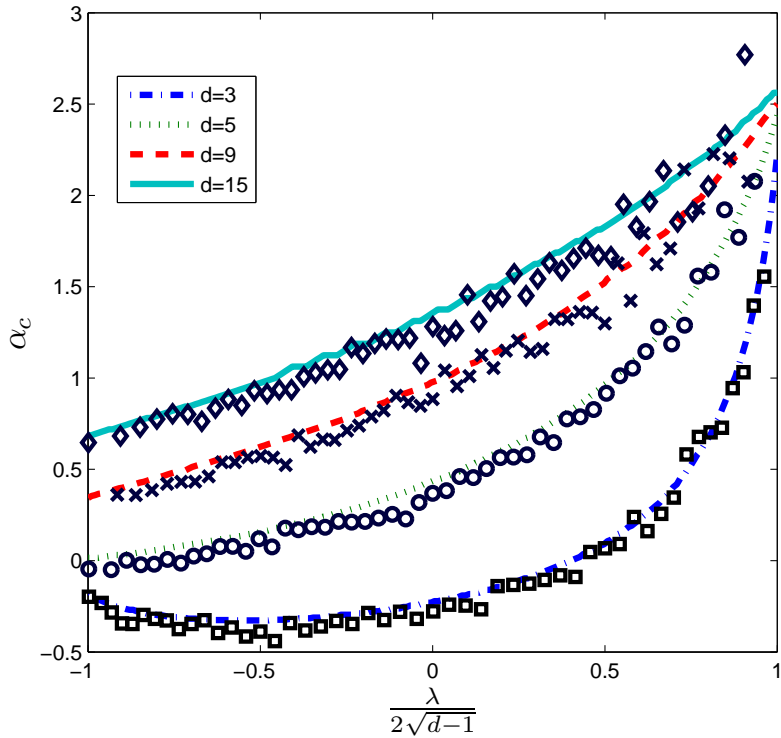}}
  \caption{\footnotesize A comparison between the value of $\alpha_c(\lambda,d)$
  for the Gaussian process $\GS$, given by equation \ref{eq:alpha_c}
  (lines) to the numerical estimation of $\alpha_c(\lambda,d)$ for
  $(n,d)$ graphs (markers).}
    \label{fig:PTNDalpha_c}
\end{figure}

The last step in the analysis is obtained by solving
(\ref{eq:alpha_c}) for  various values of $d$ and
$\lambda\in\sigma(T_d)$. The results are shown in figure
\ref{fig:PTNDalpha_c}, superimposed on the numerical data obtained
for $G(n,d)$ graphs and discussed in the previous section (figure
\ref{fig:PTNDalpha_c0}). The agreement between the $G(n,d)$ data and
the critical threshold levels computed for the $\GS$ process on
$T_d$ is perfect (including in particular the non monotonic behavior
of $\alpha_c(\lambda,d=3)$ which is reproduced as well). It strongly
supports the random waves conjecture for $G(n,d)$. As was already
stated above, the random wave conjecture is valid locally. In the
present context, however it applies globally. The proof of the
random wave conjecture is still lacking and is a challenge for
experts in probabilistic graph theory.

\section{Discussion}\label{sec:Discussion1}
The identification of a phase transition for the level sets of $\GS$
(and the observed transition for $G(n,d)$ eigenvectors) may be
considered as an extension of the percolation hypothesis for random
waves on $\mathbb{R}^2$ and for chaotic billiards
\cite{Bogomolny02}. We would like to conclude this paper by a short
comparison between the two models.

We should note that the formal derivation for both of the model is
based on the statistical properties of the corresponding random
waves model. while the heuristic justification of the percolation
hypothesis for two-dimensional random waves is based on arguments
which are not clearly justified
\cite{Foltin04,Foltin03a,Aronovitch,Keating}, theorem \ref{Thm:PT}
provides a rigorous proof to the existence of a critical level-set
for $\GS$.\\
In both of the cases, the applicability of the suggested model to
the corresponding chaotic system (a billiard or a mixing regular
graph) is a consequence of the corresponding random waves model and
not of an independent derivation. However the consistency of the
models with various numerical tests, such as
\cite{Blum,Bogomolny02,Elon07,BogomolnySle} (for billiards) and
\cite{Elon08} (for $(n,d)$ graphs) provide a firm support both for
the percolation and the random-waves models.

The arguments behind the two percolation models differ
significantly. The percolation hypothesis for two dimensional
chaotic billiards is justified in \cite{Bogomolny02,Bogomolny07} by
dimensional arguments which prevents its generalization for generic
chaotic systems, or even billiards of higher dimensions. In
particular it is based on the topological identification of a random
wave with the square lattice \cite{Bogomolny02,Keating,Foltin03a}
and the particular critical threshold $p_c=1/2$ for self-dual bond
percolation processes on the square lattice. In addition, the
neglect of correlation is based on a careful application of the
Harris criterion \cite{Harris}, which is valid only for
two-dimensional systems \cite{Bogomolny07,Foltin03a}.

Unfortunately, the identification of a critical level set for the
process $\GS$ demonstrates the same weakness, as it utilizes
repeatedly the tree structure of $T_d$. Nevertheless, the
identification of a critical level sets for the two seemingly
non-related waves ensembles hints on a universal mechanism behind
the phenomena. In addition the dependence of the critical point
(\ref{eq:alpha_c}) in the spectral parameter $\lambda$, implies that
the suggested transition for the eigenfunctions of a generic chaotic
system (assuming it indeed exists) may exhibit a more complicated
behavior then the simple model which is suggested in
\cite{Bogomolny02}.

\ack{The authors would like to thank I. Benjamini, M. Aizenman
and O. Zeitouni for enlightening discussions, comments and suggestions.\\
The work was supported by the Minerva Center for non-linear Physics,
the Einstein (Minerva) Center at the Weizmann Institute and the
Wales Institute of Mathematical and Computational Sciences) (WIMCS).
Grants from EPSRC (grant EP/G021287), ISF (grant 166/09), BSF
(710021/1) and Afeka college of engineering are acknowledged.

\noindent {\bf{Bibliography}}
\bibliographystyle{unsrt}
\bibliography{geometric1}
\end{document}